# Dynamic stabilisation of mesospin chains

Damjan Dagbjartsson[1*], Simon Banks[2] and Björgvin Hjörvarsson[1]

[1]Department of Physics and Astronomy, Uppsala University, Box 516, Uppsala, 751 20, Sweden.
[2]Department of Chemistry, King's College London, Strand, London, WC2R 2LS, United Kingdom.

*Corresponding author(s). E-mail(s): dad8@hi.is;
Contributing authors: simon.banks@kcl.ac.uk; bjorgvin.hjorvarsson@physics.uu.se;

**Abstract**

We present a rare example of a one-dimensional system with short-range range interactions for which a self-stabilizing long-range ordered phase persists to finite temperatures. Our model offers a new perspective on the origins of shape anisotropy on the mesoscopic scale in magnetic metamaterials. Specifically, we show how the combination of a physically realistic stray-field potential and the spatial dimensionality of the system can reduce the effective dimensionality of classical magnetic mesospins. The resulting emergent constraints on the magnetic excitation spectrum underpin the observed ordering, in stark contrast to the behaviour of analogous one-dimensional models with strictly isotropic spins.

## 1 Introduction

It has been one hundred years since Ising's seminal work on his eponymous model of magnetism. In the intervening period, related classical spin models have been widely studied, for their inherent interest as model magnets and as windows into phenomena across a wide range of disciplines in the natural sciences and beyond. More recently, advances in the fabrication and characterization of magnetic metamaterials have enabled the study of physical manifestations of these classical spin models with unprecedented levels of experimental control over their three primary defining characteristics: the dimensionality of the spin, the interactions between spins and the dimensionality (or the extension) of the lattice [1–4]. These magnetic structures comprise arrays of mesoscopic spins ('mesospins') – thin nanomagnets on the scale of $1-10^3$ nm with a single domain magnetic configuration over a wide temperature range, giving them characteristics of artificial classical atomic spins [5]. The thinness of the mesospins results in their magnetic easy-axes being strongly biased in-plane (a remark which, whilst generally now accepted as true and requiring no further justification, goes to the heart of the result in the current paper, as discussed below). The mesospins can be shaped lithographically, with circular discs behaving as effective XY spin analogs, and elongated islands exhibiting Ising-like behaviour [2].

Magnetic metamaterials thus facilitate the study of phase transitions, emergence, and collective dynamics with a degree of flexibility previously only available in numerical simulations. They also give insight into the impact of finite-size effects, and the significant variation in length scales over which these are relevant in different systems. However, whilst theoretical studies allow for parameters such as spin dimensionality, lattice extension/dimensionality and spin-spin interactions to be varied and combined freely, in magnetic metamaterials these parameters are subject to a complex interplay, with the significant additional consideration of the internal magnetic texture of the mesospins themselves. Of particular interest is the emergence of a self-induced anisotropy in the interactions between mesospins. This is seen to originate from a combination of the extension of the mesospins and the symmetry of the lattice defined by their arrangement[6]. Ostensibly two-dimensional square arrays have been observed to behave as sets of decoupled one-dimensional chains when suitably "dressed" by a magnetic field along the [10] direction. This behaviour hints at a new way to think of spin, and indeed lattice, dimensionality – allowing these to be seen as variable parameters in certain classes of system, with values strongly influenced both by each other and by their interdependence with the nature of the spin-spin coupling.

Thermal excitation of the pseudo-one-dimensional chains emerging in [6] results in rotation of individual mesospins. This both excites the spin-spin interactions within the chains and reintroduces coupling between the chains, significantly altering the induced inter-action anisotropy and effectively increasing the dimensionality of the lattice perceived by any given spin. It is natural to ask what effect would be seen if the second dimension of the lattice remained inaccessible even as the mesospins rotate due to thermal excitation. We explore an analogue of this question in the current paper, demonstrating that a lattice-induced interaction anisotropy can result in a one-dimensional system comprising two-dimensional (continuous) spins exhibiting behaviour reminiscent of inter-acting discrete one-dimensional components. In addition, we provide evidence that the associated model is an example of a relatively rare, physically meaningful, one-dimensional system with short-range interactions that can sustain long-range order at non-zero temperatures; the combination of lattice-induced anisotropy and vibrationnal degree of freedom, appears to overcome the usual restrictions preventing most such models exhibiting a phase transition [7–10].

## 2 Simulations

We consider a linear chain of regularly spaced circular mesospins; any given spin in the chain can be considered two-dimensional and isotropic (i.e. they are XY-like). The interactions between such mesospins in a physical system are known to be dominated by their stray fields, which exhibit an emergent anisotropy arising from a combination of lattice geometry and the mesospin internal magnetic texture[5, 6]. These closely resemble dipole-dipole interactions, but are sufficiently short-ranged to allow for only nearest-neighbour coupling to be considered significant. Due to the combination of the effective spin type and nature of the interactions, we refer to our model as "XY dipole-like" or XY-DL.

We determined the DL interaction potential using the open source micromagnetics software Mumax$^3$ [11] to model a pair of mesospins, each with diameter $D = 100$ nm and separated by a gap of $d = 40$ nm (figure 1). We take the line through the centres of the two mesospins to define the $x$-axis, with positive direction pointing from spin 1 to spin 2. Mesospin 1 forms an angle $\theta_1 = [0, 2\pi)$ with the positive $x$ direction, and similarly for mesospin 2. The calculated potential, as a function of the angles $\theta_1$ and $\theta_2$, is shown in the bottom panel of figure 1.

For comparison we have also simulated two related limiting systems: the one-dimensional XY and one-dimensional Ising models, with nearest neighbour exchange interactions in each case. The corresponding Hamiltonians are:

$$H_{\text{Ising}} = -\frac{J}{2} \sum_{\langle i,j \rangle} s_i s_j \qquad (1)$$

with $s_i \in \{-1, 1\}$ and

$$H_{\text{XY-E}} = -\frac{J}{2} \sum_{\langle i,j \rangle} \cos(\theta_i - \theta_j) \qquad (2)$$

In both cases, the sum is over all nearest-neighbour pairs of spins with boundary conditions as specified below. The notation XY-E indicates

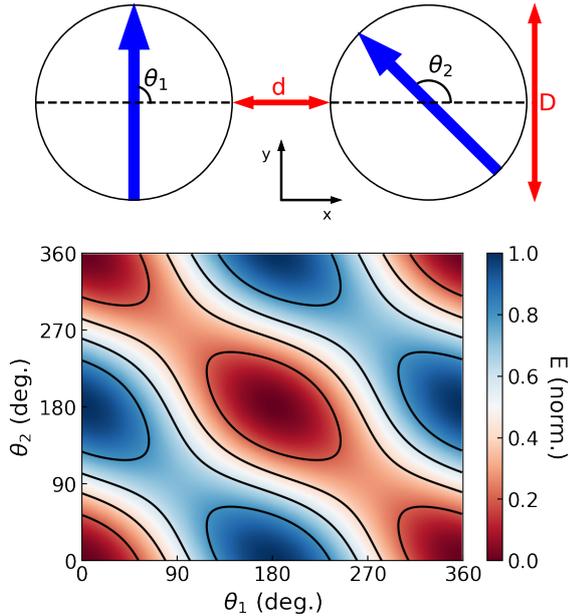

**Fig. 1 Top)** Diagram of a two-mesospin system. **Bottom)** Normalised mesospin interaction potential of the two spin system calculated using micromagnetics. The interaction is governed by stray fields and the energy is therefore dependant on the spins direction relative to the chains direction, resulting in induced anisotropy.

"XY with exchange" to distinguish this from the XY-DL model. The energy scale is defined by $J$, which is set equal to 1 throughout. For ease of comparison between these reference systems and the XY-DL mdoel, we have scaled the DL potential so that its global maximum (in dimensionless units) is equal to 1, i.e. the same pairwise maximum for both $H_{\text{Ising}}$ and $H_{\text{XY}-\text{E}}$. This effectively fixes the (unique) length of the mesospins. Tuning of the interaction strength is physically achievable by varying the size of the mesospins and the lattice parameter of the chains.

The global rotational symmetry of both $H_{\text{XY}-\text{E}}$ and $H_{\text{Ising}}$ means that the fully aligned spins in the ground state of the XY-E and Ising chains have no preferred direction in space. Importantly, the ground state of each is independent of the direction of the chain itself. One could go further and note that these models exist independently of the notion of any underlying physical arrangement of the spins (lattice). By contrast the lattice structure in the XY-DL model directly causes the anisotropy in the interaction potential, which in turn breaks the underlying rotational symmetry. Spins in the ground state are constrained to lie along the axis defined by the spin chain; the ground state is doubly-degenerate, due to the fact the spins can point in either direction along the chain, so long as they all point in the same direction.

For all models, we used a standard Metropolis Monte Carlo algorithm with $10^4$ Monte-Carlo steps per spin (MCS/s) for equilibration and $2 \times 10^6$ MSC/s for data collection at each temperature. The system was sampled after every Metropolis update. We simulated a range of system sizes, with $N$ from 10 to 10000, using both open and periodic boundary conditions. Spin updates were swept through the chain with a random new angle for each spin at each Monte Carlo step. The possible spin angles were confined to integers (360 possible angles per spin) to coincide with the grid used to calculate the potential. Whilst this technically introduces a discontinuity akin to a clock model, our tests showed no discernible impact of varying the size of this discretization provided the number of available states remained above a certain threshold. The order parameter $M$ was computed with

$$m = \sqrt{\left(\sum_{i=1}^{N} \cos(\theta_i)\right)^2 + \left(\sum_{i=1}^{N} \sin(\theta_i)\right)^2} \quad (3)$$

and averaged over each Monte Carlo step

$$M = \langle m \rangle. \quad (4)$$

## 3 Results

### 3.1 Evidence of long-range order

The behaviour of the order parameter, $M$, as a function of temperature, $T$ is shown in figure 2 for chains with $N = 30$. As expected, all three models show evidence of order when $T \to 0$. For the XY-E model, the magnetization decreases rapidly as $T$ is increased from 0, reflecting the continuous excitation spectrum. There is no evident phase transition and the apparent finite $M$ plateau at high temperatures is a known effect of finite-size (the remanent magnetization approaches zero as $N$ is increased). By contrast, the $T = 0$ order in the Ising chains persists to small finite temperatures, before the order parameter decreases through an effective phase transition at $T^*$, which we take to correspond to the point of steepest descent of $M(T)$ (or the minimum of $dM/dT$ – inset). As with the XY-E model, there is a remanent magnetization at high temperatures. This is a consequence of the finite lengths of the chains in the simulations. In the thermodynamic limit, as noted above, there is no long-range order above $T = 0$. The effect of the chain's finite size, discussed in more detail below, is to alter the energy/entropy balance in the free energy. The entropic gain associated with formation of a domain wall no-longer outweighs the corresponding energy cost; excitations gap from the ground state increases as the chains are progressively shortened.

A key result of this paper is the plot of $M(T)$ for the XY-DL model in figure 2 and again for a range of values of $N$ in figure 4. As the temperature is raised from zero, individual spins can undergo small rotations away from the ground state configuration with vanishingly small energy cost as compared to flipping of a spin in an Ising chain. The magnetization therefore shows an immediate decrease with temperature (from $M(0)$), in contrast to the finite chain Ising model for which there is a finite region in which excitations are thermally inaccessible. However, unlike the XY-E chain, the excitation energy of a given spin depends both on its direction relative to it nearest neighbours and on the angle it makes with the fixed axis defined by the underlying lattice. This suppresses spin waves at low energy and constrains excitations to small fluctuations within a largely ordered single domain for temperatures below an effective $T^* \approx 0.32$. Above this temperature, domain walls are thermally accessible and it is notable that the lowest-energy disorder-inducing excitation is a domain wall between two largely-ordered domains, rather than a long-wavelength spin wave (figure 7).

### 3.2 Finite-size and boundary effects

Our simulations of both the XY-E and Ising models show the expected effects of finite size across the full range of chain lengths studied. As the XY-E chain is lengthened, $M(T)$ falls more rapidly near $T = 0$ and the remanent high-temperature magnetization decreases, approaching zero in the thermodynamic limit (figure 3). For the Ising system, increasing the chain length leads to the expected reduction in $T^*$ proportional to $1/\ln N$ (figure 5). By contrast, whilst the XY-DL model shows some evidence of finite-size effects for very short chains, these effects not observed for chains longer than a few hundred spins (figure 5). Beyond this length, $T^*$ remains approximately constant at around 0.32 $J$, and plots of $M$ vs $T$ are essentially indistinguishable in the low temperature phase for all $N$. Whilst Peierls' argument remains valid in terms of the energy/entropy balance for the XY-DL configuration with two oppositely oriented perfectly-ordered domains, this arrangement has a higher overall free energy than thermally accessible excitations of a single domain configuration. The net result is a region of apparent long-range order, undergoing a phase transition at $T^*$. Figure 6 shows a comparison of this finite-size behaviour in simulations of the XY-DL model when using open vs periodic boundary conditions. For short chains ($N < 80$) with periodic boundary conditions, the finite-size effects are reminiscent of Ising-like behaviour. As might be expected, simulations using open boundaries show an increasing influence of "looser" edge spins as the chain length is decreased. With edge spins having just one neighbour, they are more susceptible to thermal fluctuations; this manifests as a reduction in the stability of the ordered phase and an initial reduction in $T^*$ as we move towards shorter chain lengths. As the chain length is further shortened, additional finite-size effects first contribute to, and then dominate, the ordering mechanism. The impact of the edge spins thus diminishes and $T^*$ increases.

## 4 Discussion

It is often assumed that one-dimensional systems with short-range interactions possess no thermodynamically stable phase with long-range order above $T = 0$ in the thermodynamic limit $N \to \infty$ (where $N$ is the number of spins). In the case of continuous spins, this assertion is generally supported by reference to the Mermin-Wagner theorem[7], in which long-wavelength spin waves are energetically accessible for all non-zero temperatures for sufficiently large systems. For systems of discrete spins, the assumed general absence of long-range order is typically explained with reference to the heuristic arguments of Peierls, famously explained by Landau and Lifshitz, regarding the balance of energetic and entropic contributions to the change in free energy on formation of a domain wall. In reality the picture is more complex and broad statements regarding the absence of long-range order in one-dimensional models typically assume certain (at times, unspecified) additional constraints. For example, Mermin and Wagner were clear that their arguments applied to isotropic continuous spins – leaving the questions of whether, and to what extent, anisotropy might facilitate long-range order to other studies. Likewise, it has long been established that one-dimensional systems with short-range interactions can exhibit long-range order in the presence of external fields or spin inhomogeneity. Cuesta and Sánchez provided a valuable addition to the literature on this topic, with their development of a general theorem for the conditions leading to the non-existence of phase transitions in one-dimensional systems with short-range interactions. A notable result of their work was to highlight examples of physically meaningful transitions in one-dimensional models for which the conditions of their general theorem were not met. Even for systems for which no long-range order is strictly stable, there remains the possibility that topological order might be viable, perhaps most famously exemplified in two dimensions by the 2dXY model, which undergoes a Kosterlitz-Thouless-Berezinskii (KTB) transition from a paramagnetic to a low-temperature critical phase. Such transitions are seen in a variety of physically interesting systems, both at surfaces and in monolayers[12], including in mesoscopic spins.

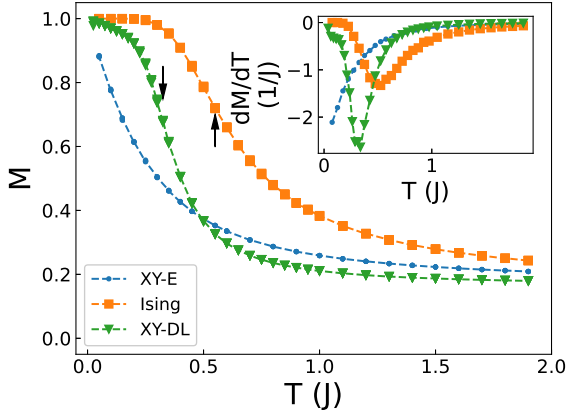

**Fig. 2** The effect of spin dimensionality and interaction potential on ordering of finite chains. M vs T of Ising, XY-E and XY-DL chains at $N = 30$. From the derivatives (inset), an inflection point, $T^*$, can be seen in the Ising and XY-DL chains but is absent in the XY-E chain. $T^*$ is also marked with arrows on the M vs T graph.

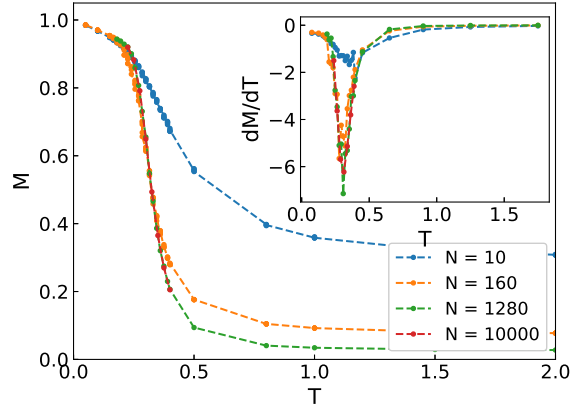

**Fig. 4** The effect of system size. M vs T of a XY-DL chain with lengths $N = 10, 20, 160, 1280$. The inset shows the first derivative of the magnetisation. The profile at low $T$ appears to converge with $N$ with $T^* \approx 0.32$.

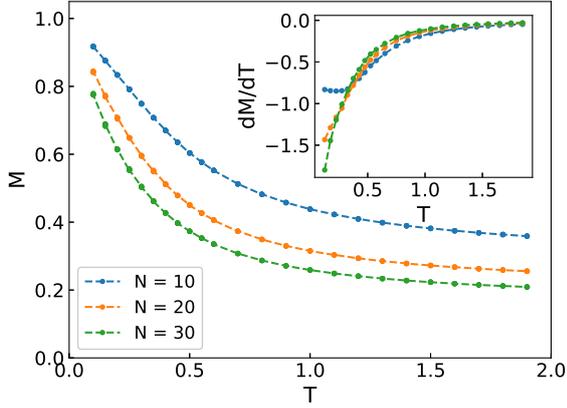

**Fig. 3** The effect of system size. M vs T of a XY-E chain with lengths $N = 10, 20, 30$. The inset shows the first derivative of the magnetisation. The order parameter, due to the finite size of the chains, decreases immediately and fast as $T > 0$. As the chains length increases, $M$ decreases even faster, which can be seen by the decrease in the derivative at $T \to 0$ as $N$ increases.

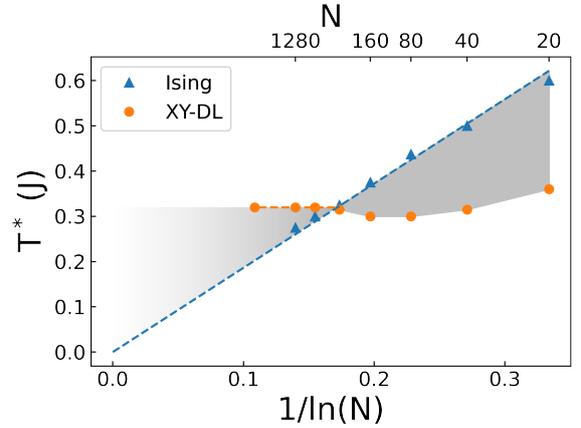

**Fig. 5** The effect of spin dimensionality on order. $T^*$ vs $1/\ln(N)$ for Ising and XY-DL chains with open boundaries. The dotted lines are fits and the shaded areas highlight the difference in the order between the two systems.

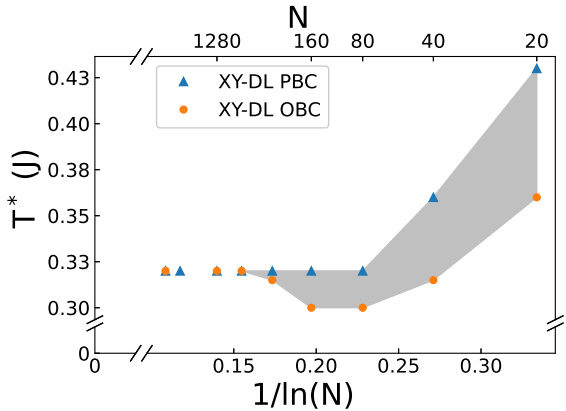

**Fig. 6** The effect of open and periodic boundaries. T* vs $1/\ln(N)$ for XY-DL chains with open(OBC) and periodic(PBC) boundary conditions. The decrease in order due to the edge effect is highlighted.

## 5 Conclusion

We have shown how a realistic stray-field potential coupling classical magnetic mesospins in a one-dimensional chain effectively reduces the dimensionality of the mesospins in the ground state, constraining them to lie along a lattice-defined easy axis. This emergent effective spin anisotropy suppresses long-wavelength excitations to the extent that the system can support a stable phase with long-range order at finite temperatures. The scaling of the effective transition temperature with system size exhibits Ising like behaviour for very small systems; however, it rapidly converges to $T^*/J \approx 0.32$, a value that appears to be independent of system size for all $N > 80$.

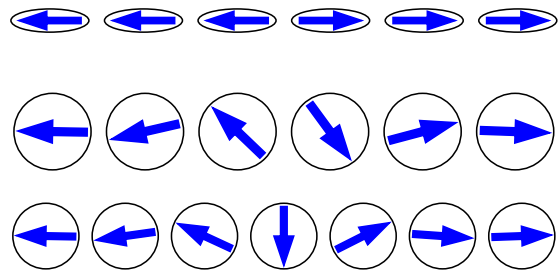

**Fig. 7** Domain walls calculated by relaxing a two domain Monte Carlo simulation. The chains shown here have been truncated from a $N = 20$ simulation. **Top)** Ising chain with domain boundary energy cost of $\Delta E_{DW} = 2J$. **Center)** Asymmetrical XY-DL chain with domain boundary energy cost of $\Delta E_{DW} \approx 1.8J$. **Bottom)** Symmetrical XY-DL chain with domain boundary energy cost of $\Delta E_{DW} \approx 1.8J$

This result is important for two reasons. Firstly, our model represents a rare example of a physically meaningful one-dimensional system with short-range interactions that can support long-range order above $T = 0$. This stability arises from the interplay of the spatial dimensionality and inter-spin coupling, and is in stark contrast to the behaviour observed for exchange coupled Ising or XY spins in one or two dimensions. Secondly, our results provide a new perspective on the origins of shape anisotropy at the mesoscopic scale. In a physical realisation of our model, the overall demagnetizing field of the system is not inherently determined by any one of the spin dimensionality, the lattice dimensionality, nor the inter-spin coupling. Instead, the emergent shape anisotropy results from the interplay between the dominant coupling potential and spatial dimensionality of the system. The result is a self-induced easy axis in the constituent mesospins, significantly constraining the thermally accessible excitations and dominating the equilibrium magnetic behaviour at small finite temperatures.

We suggest that our results are immediately applicable in the context of the design and characterization of low-dimensional magnetic metamaterials, and may have implications for the understanding of the origins of shape anisotropy more broadly.